\documentstyle[prb,aps,multicol,epsf]{revtex}
\begin{document}
\tolerance 10000
\def\binom#1#2{{#1\choose#2}}
\tighten
\draft
\title{Interplay of Mott Transition and Ferromagnetism in
the Orbitally Degenerate Hubbard Model}
\author{Raymond Fr\'esard$^{1,2}$ 
\thanks{Present address: Institut de
Physique, Universit\'e de Neuch\^atel, A.-L. Breguet 1, CH-2000
Neuch\^atel} 
and Gabriel Kotliar$^2$}
\vspace*{0.5truecm}
\address{
$^1$ Physics Department, Shimane University, Nishikawatsu-cho 1060,\\
Matsue 690, Shimane, Japan\\
$^2$ Department of Physics and Astronomy, Rutgers University,
Piscataway, NJ-08855-0849, USA}
\date{\today}
\maketitle
\widetext
\begin{abstract}
\begin{center}
\parbox{14cm}{A slave boson representation for the degenerate Hubbard model
is introduced. The location of the metal to insulator transition that occurs
at commensurate densities is shown to depend weakly on the band degeneracy $M$.
The relative weights of the Hubbard sub-bands depend strongly on $M$, as well
as the magnetic properties. It is also shown that a sizable Hund's rule 
coupling is required in order to have a ferromagnetic instability
appearing. 
The metal to 
insulator transition driven by an increase in temperature is a strong function
of it.}
\end{center}
\end{abstract}

\pacs{
\hspace{1.9cm}
}
\begin{multicols}{2}
There has been dramatic progress in our understanding
of the Mott transition  in the last few years.
Careful experimental studies of systems in
the vicinity of the Mott transition have been carried out
\cite{Tokura} and two new theoretical tools, slave bosons
mean field theories (see for instance \cite{FW} and references
therein), and the limit of infinite dimensions
have been adapted to its study.
For a review see \cite{Georges}.
Most of the modern  work has focused on the {\it single
band Hubbard model}. Now that both the metallic and the Mott
insulating states of the (doped) titanates and vanadiates have been
studied experimentally \cite{Tokura,Goodenough} (corresponding to $3d1$ and 
$3d2$ configurations in the Mott insulating state), there is a need for a
theoretical framework allowing for understanding the Mott transition
for arbitrary degeneracy and density. This paper is aimed to provide
such a technique and to apply it to a variety of quantities that
cannot be obtained easily using alternative approaches. Most of the results
are obtained in a closed analytical form, allowing for a qualitative 
understanding of the physical situation.

In this work we investigate the effect of strong Coulomb interaction
in systems with {\it orbital degeneracy}.
Such a
situation is realized in virtually all  transition metals and transition metal
oxides.
These systems contain d electrons in cubic or trigonal environments,
the crystal field can only   lift partially
the  degeneracy of the d-bands,
down to 2  as is the case of  $V_2 O_3$
\cite{Ranninger} or  3 as in  $La Ti O_3$.
Our goal is to understand how degeneracy affects the behavior
of the different physical  quantities near the Mott transition.
To carry out the investigation we
extend the slave boson technique which has been very successful in the
study of the Mott transition, to the orbitally degenerate case. 
Compared to the variational wave function approach \cite{Gutzwiller}
\cite{Lu}
our formalism  is  more flexible since, as we demonstrate
in this paper,  it allows  us to calculate
a variety of quantities which are not easily accessible
to the variational approach,
as a function of 
the correlation strength and doping.
It can  also be improved systematically by performing a loop
expansion around the saddle point. 
Our main results are the following :a) low energy single particle
quantities such as the critical value of the interaction strength
at which the transition occurs,  the
quasiparticle residue and the single particle Mott
Hubbard  gap depend very weekly
on degeneracy justifying the agreement between theory and experiment
when it was applied to orbitally degenerate systems. b)
the relative weights  of the Hubbard bands depend strongly
on degeneracy in agreement with other methods. \cite{Eskes}
c) the degeneracy temperature decreases with increasing band degeneracy.
d) the magnetic properties, the magnetic susceptibility  and its
associated Landau parameter in the paramagnetic phase   and the
magnetic  phase diagram is strongly
modified from the one band case. 

The  Hamiltonian  describing the low-energy properties of these systems 
is commonly written as: 
\begin{eqnarray}
H &=& \sum_{i,j,\sigma,\rho} t_{i,j} c^+_{i,\rho,\sigma}
c_{j,\rho,\sigma}
+ U_3 \sum_{i,\rho} n_{i,\rho,\uparrow}n_{i,\rho,\downarrow}\nonumber\\
&+&U_1\sum_{i,\rho'\neq\rho}n_{i,\rho,\uparrow}n_{i,\rho',\downarrow}
+U \sum_{i,\sigma,\rho'<\rho}n_{i,\rho,\sigma} n_{i,\rho',\sigma}
\end{eqnarray}
where $\sigma$ is a spin index for the up and down
states while $\rho$ is labeling the M bands, and $U_n \equiv U+nJ$. 
Taking $J$ finite accounts for the Hund's rule
coupling.

As for any model with on-site interaction, a slave boson
representation can be introduced, mapping all the degrees of freedom onto
bosons. We can re-write any atomic state with
the help of a set of pseudo-fermions $\{f_{\alpha}\}$ and slave
bosons $\{\psi_{\alpha_1,... \alpha_m}^{(m)}\}$ ($0\leq m\leq 2M$). 
$\psi_{\alpha_1,... \alpha_m}^{(m)}$ is the slave boson associated 
with the atomic
state consisting of $m$ electrons in states $| \alpha_1,...,\alpha_m>$
where $\alpha$ is a composite spin and band index.
By construction it is symmetric under any permutation of 2 indices, and 0
if  any   2 indices are equal.
We can now write the creation operator of a physical 
electron in terms of the slave particles as:
\begin{equation} \label{eqc}
c^+_{\alpha} = \tilde{z}^+_{\alpha} f^+_{\alpha}
\end{equation}
$\tilde{z}^+_{\alpha}$
describes the change in the boson occupation numbers when an
electron in state $\alpha$ is created as:
\begin{equation} \label{eqzt}
\tilde{z}^+_{\alpha} = \sum_{m=1}^{2M} \sum_{\alpha_1<.<\alpha_{m-1}} 
\psi^{+ (m)}_{\alpha,\alpha_1,...,\alpha_{m-1}}
\psi^{(m-1)}_{\alpha_1,...,\alpha_{m-1}} \quad \alpha_i \neq \alpha
\end{equation}
The 
operators $\tilde{z}^+_{\alpha}$ in Eq. (\ref{eqzt}) describes
the change in the slave boson occupation as a many channel
process. In order to recover the correct 
non-interacting limit at mean-field level, one has to observe that the
classical probability for these processes to happen is
not simply given by taking the Bose fields in (\ref{eqzt}) to be
given by their classical values, but 
by introducing normalization factors $L_{\alpha}$ and $R_{\alpha}$
\cite{Kotliar,FW} as 
$z^+_{\alpha} = \sum \psi^{+ (m)} L_{\alpha} R_{\alpha} \psi^{(m-1)}$, where:
\begin{eqnarray}
R_{\alpha} & = & [1-\sum_{m=0}^{2M-1} \sum_{\alpha_1,<.<,\alpha_m} 
\psi^{+ (m)}_{\alpha_1,.,\alpha_{m}} 
\psi^{(m)}_{\alpha_1,.,\alpha_{m}}
]^{-\frac{1}{2}} \quad \alpha_i\neq \alpha \nonumber \\
L_{\alpha} & = & [1-
\sum_{m=1}^{2M} \sum_{\alpha_1<.<\alpha_{m-1}} 
\psi^{+ (m)}_{\alpha,\alpha_1,.,\alpha_{m-1}} 
\psi^{(m)}_{\alpha,\alpha_1,.,\alpha_{m-1}}]^{-\frac{1}{2}} \quad .
\end{eqnarray}
Namely $L_{\alpha}$ normalizes to 1 the probability that no electron in
state $|\alpha>$ is present on a site before one such electron hops to that
particular site, and $R_{\alpha}$ makes sure that it happened. Clearly the
eigenvalues of the operators $L_{\alpha}$ and $R_{\alpha}$ are 1 in the 
physical subspace.
Now, the redundant degrees of freedom are projected out with the
constraints:
\begin{eqnarray}
f^+_{\alpha} f_{\alpha} - &\sum_{m=1}^{2M}& \sum_{\alpha_1<.<\alpha_{m-1}} 
\psi^{+ (m)}_{\alpha,\alpha_1,.,\alpha_{m-1}} 
\psi^{(m)}_{\alpha,\alpha_1,.,\alpha_{m-1}} = 0 
\nonumber \\
&\sum_{m=0}^{2M} &\sum_{\alpha_1<.<\alpha_{m}} 
\psi^{+ (m)}_{\alpha_1,.,\alpha_{m}} 
\psi^{(m)}_{\alpha_1,.,\alpha_{m}} - 1 = 0 \quad.
\end{eqnarray}
We obtain the Lagrangian at $J=0$ as:
\begin{eqnarray}
&L&= \sum_{i \alpha}f^+_{i,\alpha} (\partial_{\tau} - \mu + i
\lambda_{i,\alpha})f_{i,\alpha} - i\Lambda_{i} + \sum_{i,m} 
\sum_{i \alpha_1<.<\alpha_m}
\nonumber \\
&    
\psi&^{+ (m)}_{i,\alpha_1,.,\alpha_m} ( \partial_{\tau} + i\Lambda_{i}
+ U \binom{m}{2} - i \sum_{j=1}^{m} \lambda_{i,\alpha_j})
\psi^{(m)}_{i,\alpha_1,.,\alpha_m} \nonumber \\
&+& \sum_{i,j,\alpha} t_{i,j} z^+_{i, \alpha} 
f^+_{i,\alpha } z_{j, \alpha} f_{j, m, \alpha} \quad .
\end{eqnarray}

\begin{figure}
\narrowtext
\centerline{\epsfxsize=7cm
\epsfbox{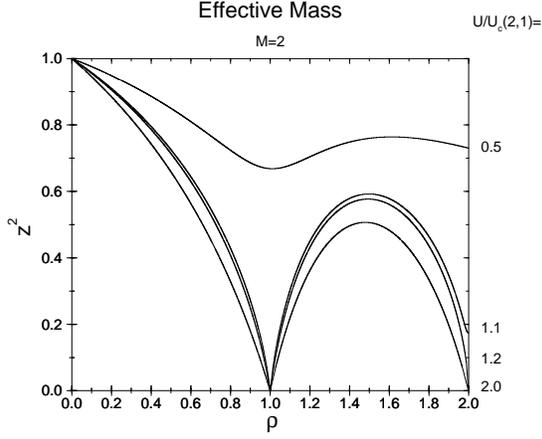}}
\vspace{5pt}
\caption{Inverse effective mass in the 2 band model
as a function of density for several
values of $U$.}
\label{fig1}
\end{figure}

We now proceed to the mean-field theory, and we investigate the
paramagnetic, paraorbital saddle-point. The latter is obtained after
integrating out the fermions, and setting all bosonic fields to their
classical value. 
The Mott transition that occurs at commensurate density $n$ is best
discussed by projecting out occupancies that are larger than $n+1$ and
smaller than $n-1$ (if any). 
The constraints allows
for eliminating the variables $\psi^{(n-1)}$ and $\psi^{(n)} $
to 
obtain the grand potential
at $n$: 
\begin{eqnarray}\label{sp0}
\Omega(D) &=& (1-2D^2)D^2(\sqrt{b_{n,M}} + \sqrt{c_n})^2 \epsilon_0 
\nonumber \\
&+& U(D^2 +\binom{n}{2})- \mu \rho
\end{eqnarray}
with $\epsilon_0 \equiv 2M\int d\epsilon \epsilon \rho(\epsilon)
f_F(z^2\epsilon +\lambda_0 -\mu)$, $D^2 \equiv
\binom{2M}{n+1}
\psi^{(n+1) 2}$, $b_{n,M} \equiv (2M-n+1)/(2M-n)$, and $c_n \equiv (n+1)/n$. 
Minimizing Eq. (\ref{sp0}) with respect
to $D$ yields
a critical
interaction strength at which $D$ vanishes. 
It reads $U_c^{(n,M)} =
-\epsilon_0(\sqrt{b_{n,M}} + \sqrt{c_n})^2$
which reproduces the results of  
the Gutzwiller approximation 
\cite{Gutzwiller,Lu}. 
This locates the Mott transition.  Restricting
ourselves to a
flat density of states 
we can relate the critical interaction strength to the
band width $W$. We obtain:
\begin{equation}\label{cri2}
U_c^{(n,M)} = \frac{nW}{4M}(2M-n) (\sqrt{b_{n,M}} + \sqrt{c_n})^2
\end{equation}
Its band degeneracy dependence is
fairly weak \cite{Lu}.
The effective mass of the quasi-particles diverges at the Mott
transition. We obtain:
\begin{equation}
\frac{m}{m^*} = z^2 = \frac{(\sqrt{b_{n,M}} + \sqrt{c_n})^2}{8} 
\quad \frac{U_c^{(n,M) 2}-U^2} {U_c^{(n,M) 2}}
\end{equation}
Due to the particular form of the coefficients $b$ and $c$ the
dependence on the band degeneracy 
is weak. 
The critical interaction 
strength increases with $M$ so the quasi-particle residue $Z$ increases 
slightly  with M.
For small values of U,  (which we treated without 
projecting out higher occupancies), Z decreases with
increasing $M$. So there is a  crossover value of the
interaction strength  beyond which the system becomes more metallic with
increasing  $M$  \cite{RFGK}. 
As a function of the hole doping $\delta$, the quasi-particle residue
vanishes for $\delta$ going to 0 above $U_c^{(n,M)}$ as:
\begin{eqnarray}
z^2 &=& \frac{\delta}{2}(b_{n,M} - c_n) + \frac{|\delta|}{2}
((b_{n,M} + c_n)\nonumber\\
&\times& \sqrt{1+4\varphi_{n,M}} + 4 \sqrt{b_{n,M} c_n \varphi_{n,M}})
\end{eqnarray}
where we introduced:
\begin{equation}
\varphi_{n,M} \equiv \frac{ U_c^{(n,M) 2}
\frac{b_{n,M}c_n}{(\sqrt{b_{n,M}} +
\sqrt{c_n})^4}}
{(U-U_c^{(n,M)})(U-U_c^{(n,M)}
(\frac{\sqrt{b_{n,M}}-\sqrt{c_n}}{\sqrt{b_{n,M}}+\sqrt{c_n}})^2)}
\end{equation}
The expression of the quasi-particle residue consists of 2 contributions 
which are either symmetric or antisymmetric with respect to particle or hole 
doping. The antisymmetric contribution vanishes for $n=M$ as a consequence
of the particle-hole symmetry. 
The asymmetry of $z^2$ on particle or hole doping
is seen to increase under an increase of $|n-M|$. It vanishes more slowly 
for hole doping (for $n\le M$) than for
particle doping, for increasing degeneracy at fixed $n$, for
increasing degeneracy  at $n=M$ and under an increase of $U$. As an example 
we calculate the effective mass for the 2 band
model and show it on Fig. \ref{fig1}, which has been 
calculated without projecting out higher occupancies.

Interestingly
we also obtain a Mott gap. Indeed the number of quasi-particles is a
continuous function of their chemical potential
$\mu-\lambda_0/2$. However the saddle-point equations show that the
Lagrange multiplier $\Lambda$ jumps when going through the Mott
gap which implies that $\lambda$ is going to jump as well, and so does
$\mu$. As a result we obtain the Mott gap $\Delta \equiv
\lim_{\delta \rightarrow 0^-}\mu(\delta) 
-\lim_{\delta \rightarrow0^+}\mu(\delta) $ as:
\begin{equation}\label{eqgap}
\Delta = \sqrt{(U-U_c^{(n,M)})(U-U_c^{(n,M)}
(\frac{\sqrt{b_{n,M}}-\sqrt{c_n}}{\sqrt{b_{n,M}}+\sqrt{c_n}})^2)}
\end{equation}
\begin{figure}
\narrowtext
\centerline{\epsfxsize=7cm
\epsfbox{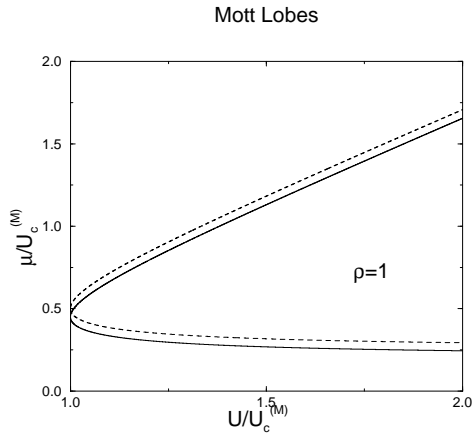}}
\vspace{5pt}
\caption{Chemical potential for $n=1$ for the 1 band (dashed line)
and 2 band (full line) models.}
\label{fig2}
\end{figure}
In the limit of $U>>U_c^{(n,M)}$, the Mott gap is given by $U$, while it
closes at $U_c^{(n,M)}$ as $\Delta \sim U_c^{(n,M)}\sqrt{U/U_c^{(n,M)}-1}$,
the square root behavior being typical of slave boson
mean-field theories. It can be read
from Fig. \ref{fig2} where it is compared to the 1 band case as obtained
by Lavagna \cite{Lavagna}. 
Clearly going from 1 band to 2 bands does not imply
a big difference in the Mott gap. Indeed we obtain the
$\Delta/U_c^{(n,M)}$ is independent of $M$ at $n=M$, while for fixed
$n$ it depends very weakly on $M$. 

\begin{figure}
\narrowtext
\centerline{\epsfxsize=7cm
\epsfbox{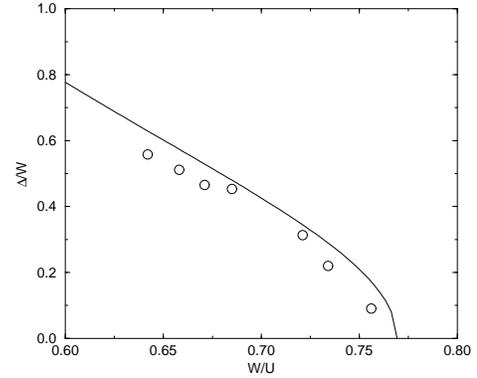}}
\vspace{5pt}
\caption{Dependence of the Mott gap on the band width for $n=1$ and $M=3$. 
Circles:
experimental data of Okimoto {\it et al}$^{12}$.}
\label{figexp}
\end{figure}

Our result can be compared to experimental data. For the series
$La_xY_{1-x}TiO_3$, Okimoto {\it et al} \cite{Okimoto} measured how
the gap depends on the band width. Assuming (for large ratio $U/W$)
$\Delta \sim U-W$ we obtain out of their data $U = 3.2 eV$. Inserting
this and the experimental value of $(U/W)_c \sim 1.3$ in Eq.(\ref{eqgap}) we 
can compute $\Delta/W$ as a function of $W/U$ and compare it with
experiment on fig. \ref{figexp}. The experimental trend is clearly
reproduced and the quantitative agreement is very satisfactory.

We now turn to the Hund's rule coupling dependence and treat as an example
the two band model around the $n=1$ Mott insulating lobe. At $\rho=1$
the grand potential at the saddle-point reads:
\begin{eqnarray} \label{mfeq2}
\Omega &=& \frac{4}{3} \epsilon_0 
\left(1-2r^2 
\right)
(r+ 
(d_0 +d_x + \Delta_0)/ \sqrt{2}
)^2 \nonumber \\ 
&+& (U+3J)\Delta_0^2 + (U+J) d_x^2 
+ Ud_0^2 - \mu \rho \quad .
\end{eqnarray}
with $d_0 \equiv (\psi^{(2)}_{\uparrow,\uparrow} + 
\psi^{(2)}_{\downarrow,\downarrow})/\sqrt{2}$, $d_x \equiv 
(\psi^{(2)}_{\uparrow,\downarrow} + 
\psi^{(2)}_{\downarrow,\uparrow})/\sqrt{2}$, $\Delta_0 \equiv 
(\psi^{(2)}_{\uparrow\downarrow,0} + 
\psi^{(2)}_{0,\downarrow\uparrow})/\sqrt{2}$, 
$r^2 \equiv d_0^2 +d_x^2 +\Delta_0^2$
and $\lambda \equiv
\sum_{\alpha}  \lambda_{\alpha}/2$, and we have used the constraints
to remove the variables $\psi^{(0)}$ and $\psi^{(1)}$.

Such an expression differs from an ordinary Ginzburg-Landau free
energy in that respect that it cannot be written as a 4th order
polynomial in the variables $d_0$, $d_x$ and $\Delta_0$. As a result,
if there were to be a critical point for 1 field, it would be critical
for the other ones as well. 
We obtain the location of the Mott transition as:
\begin{equation}
U^{(2)}_{c,(J)} = U^{(2)}_{c,(0)} ( 1 - \frac{4}{3} \quad
\frac{J}{U} + O (J/U)^2) \quad.
\end{equation}
Another regime of interest is the large $J$ regime. 
There we obtain the location of the Mott transition as:
\begin{equation}
U_c^{(2)} = -\frac{2}{3} \epsilon_0(3+2\sqrt{2})(1-\frac{8}{9} \frac{\epsilon_0}{J}) 
+O((\frac{\epsilon_0}{J})^2)
\end{equation}
and thus decreasing $J$ from $\infty$ leads to an increase of the critical 
interaction. Another intriguing feature of transition metal oxides such
as $V_2O_3$, is the metal to insulator transition that occurs in the vicinity
of the tri-critical point under an increase of temperature. It has recently
been interpretated \cite{Rozenberg} as the transition from a Fermi liquid
with finite quasi-particle residue $Z$ to an insulator with $Z=0$. In other 
words there is a finite coherence temperature $T_{coh}$ at which the coherence
of the Fermi liquid (and $Z$) vanishes. This result was obtained in the 
dynamical mean-field approximation to the 1 band model, which becomes exact 
in the limit of large dimensions, and recovered in the Gutzwiller Approximation
\cite{Doll}. 
At finite $T$ there is a first order metal to insulator transition at
a critical $U_c^{(M)}(T)$:
\begin{equation}
U_{c}^{(M)}(T) = U_c^{(M)}(0) - \sqrt{8 U_c^{(M)}(0) T \ln{2M}}
\end{equation}
Thus an increase in 
temperature may produce a metal to insulator transition, which is
consistent with the experimental situation in $V_2O_3$. In the
dynamical mean-field approximation at finite temperatures
there is an interaction
strength
$U_{c 2}(T)$ at which  
the metallic solution ceases to exist.
This quantity can also be evaluated
in our slave boson scheme and is given by: 
\begin{equation}
U_{c 2}^{(M)}(T) = U_c^{(M)}(0) 
(1 - \alpha_M (T/W)^{\frac{2}{3}})
\end{equation}
with $\alpha_1 \sim 2.53$ and $\alpha_2 \sim 3.32$.

\begin{figure}
\narrowtext
\centerline{\epsfxsize=7cm
\epsfbox{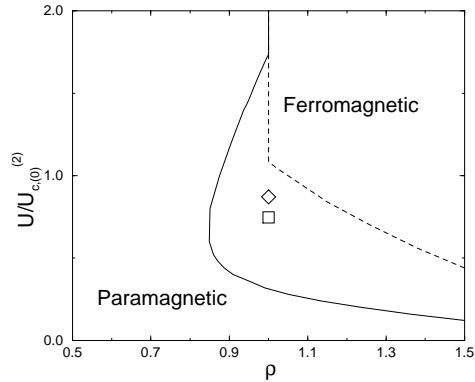}}
\vspace{5pt}
\caption{Instability line of the paramagnetic phase for $U/J=10$ (dashed line)
and $U/J=5$ (solid line). The diamond (square) indicates the position of the 
Mott transition for $U/J=10$ ($U/J=5$).}
\label{fig4}
\end{figure}

\begin{figure}
\narrowtext
\centerline{\epsfxsize=7cm
\epsfbox{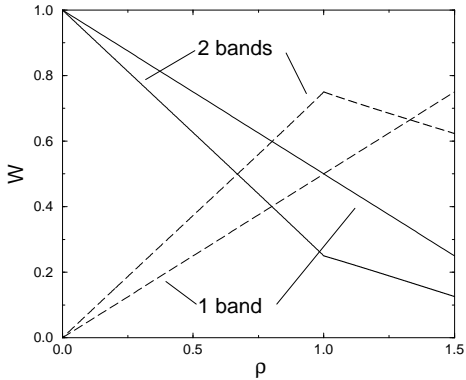}}
\vspace{5pt}
\caption{Spectral weight of the upper (dashed lines) and lower (solid
lines) Hubbard bands for the 1 band
and 2 band models, at $U = 2 U_c^{(M)}$.}
\label{fig5}
\end{figure}

We now turn to the calculation of the magnetic susceptibility. Here we 
generalize the calculation of Li {\it et al} \cite{Li91} to the 2 band 
model. The
linear response to an external magnetic field is obtained as a 1 loop 
calculation of the correlation function of the slave boson fields in the
spin-antisymmetric band-symmetric channel. 3 fields couple in this channel
\cite{RFGK}: $\chi_- \equiv \frac{1}{2}\sum_{\alpha} \sigma 
\psi^{(1)}_{\alpha}$,
$\chi_+ \equiv \frac{1}{\sqrt{2}}
(\psi^{(2)}_{\uparrow,\uparrow}-\psi^{(2)}_{\downarrow,\downarrow})$ and
$\kappa \equiv \frac{1}{2}\sum_{\alpha} \sigma \lambda_{\alpha}$, and
the magnetization is expressed in terms of slave bosons as $M = 4 d_0 \chi_+
+ 2 \psi^{(1)} \chi_-$.
The resulting susceptibility arises as an RPA form \cite{RFGK}
\begin{equation}
\chi_S = \frac{\chi_0}{1+F_0^a \chi_0/N(E_F)} \quad .
\end{equation}
We now determine the instability line of the 
paramagnetic phase with respect to ferromagnetism. For $J=0$ we find
no ferromagnetic instability
even near the Mott transition, while for finite $J$ we find that the
Mott metal insulator transition may be preempted by the appearance of
a ferromagnetic phase. In other words, a sufficiently strong
Hund's rule coupling turns a Mott insulator into a ferromagnet. Originally
the Hubbard model was introduced in order to describe ferromagnetism
in narrow band systems, but it has been recently established 
that the ground state is not ferromagnetic for any
reasonable values of the parameters 
on generic lattices \cite{Moeller,Wurth}. We find that the 
ground state is much more likely to be ferromagnetic in the degenerate model
for finite $J$, as shown in Fig. \ref{fig4}.
Our method can be applied to the calculation of dynamical quantities too. 
In the strong coupling regime the one-particle excitation spectrum is split
off into several pieces, each of them carrying some fraction of the spectral
weight (which are adding up to 1 so as to fulfill the sum rule). The various 
pieces are following from the discrete atomic levels, which are well
separated by multiples of $U$, broadened by exchange processes. Let us now 
determine the fraction of the spectral weight carried by each sub-band.
In our language the low energy excitations are involving the 
field $\psi^{(1)}$,
and the high energy excitations centered around $U$ the field $\psi^{(2)}$. 
Higher energy excitations involving higher local occupancies are left out.
We obtain the spectral 
weights in both
bands from the decomposition of the physical electron operator 
(Eqs. (\ref{eqc},\ref{eqzt})).
Accordingly the spectral weight of the
Green's function $T <c_{\alpha}(\tau) c^+_{\alpha}(0)>$ in the
lower Hubbard band ($W_{LHB}$) and the upper Hubbard band ($W_{UHB}$) are given
by:
\begin{eqnarray}
W_{LHB} &=& <\psi^{+ (0)}\psi^{(0)} + \psi^{+ (1)}_{\alpha} 
\psi^{(1)}_{\alpha} > \nonumber \\
W_{UHB} &=& \sum_{\beta\neq\alpha} <\psi^{+ (1)}_{\beta} \psi^{(1)}_{\beta} 
+ \psi^{+ (2)}_{\alpha\beta} \psi^{(2)}_{\alpha \beta}>
\end{eqnarray}
and are shown in fig. \ref{fig5}. Here the weights do not quite add up to
1 in the 2 band case because we projected out occupations larger than 2.
In other words, on top of the 2 sub-bands which are considered here, there
appear a second upper Hubbard band (centered around $3U-3\mu$, corresponding 
to triple occupancy) which
is becoming relevant in the particle doped regime. To a very good accuracy 
its contribution to
the spectral weight is given by $1-W_{LHB}-W_{UHB}$.
Clearly the degeneracy plays an important role as the weight of the upper
band at $n=1$ in the strong coupling regime is given by $(2M-1)/2M$. 

In summary we introduced a slave boson representation of the degenerate
Hubbard model. We obtained that the band degeneracy has a weak influence on
the location of the Mott transition, while the degeneracy temperature and the
dynamical and magnetic properties strongly depend on it. We also showed
that no ferromagnetic instability occurs unless the  Hund's rule coupling 
becomes sizable, yielding a generic scenario for ferromagnetism in
transition metals and transition metal oxides. In that case a
ferromagnetic
instability may even shadow the Mott transition. RF
gratefully acknowledges financial support from the Fonds National Suisse de
la Recherche Scientifique, as well as Rutgers University and Neuch\^atel
University for hospitality were part of this work has been done. This work
has been partially supported by the NSF under grant DMR 95-29138.
\par

\end{multicols}


\begin{thebibliography}{52}
%
\bibitem{Tokura} Y. Tokura {\it et al}, Phys. Rev. 
Lett. {\bf 70}, 2126 (1993).
\bibitem{FW}R. Fr\'esard  and P. W\"olfle,
{\em Proc. of the Adriatico Research Conference and Miniworkshop Strongly
Correlated Electron Systems III}, Eds. G. Baskaran {\it et al},
Int. J.
of Mod. Phys. B {\bf 6}, 685 (1992); Erratum, Int. J. Mod. Phys. B
{\bf 6}, 3087 (1992).
\bibitem{Georges} A. Georges {\it et al}, Rev. Mod. Phys. {\bf 68}, 13
(1996).
\bibitem{Goodenough} H.C. Nguyen and J.B. Goodenough, Phys. Rev. B
{\bf 52}, 8776 (1995); F. Inaba {\it et al}, Phys. Rev. B
{\bf 52}, R2221 (1995)
\bibitem{Ranninger} C. Castellani {\it et al},
Phys. Rev. B {\bf 18}, 4945 (1978).
\bibitem{Eskes} H. Eskes  {\it et al}, Phys. Rev. B {\bf 50}, 17980 (1994).
\bibitem{Gutzwiller} K.A. Chao and M. Gutzwiller, J. Appl. Phys.~{\bf 42},
1420 (1971).
K.A. Chao J. Pys. C {\bf 7}, 127 (1974); J. B\"unemann and W. Weber, Preprint
Sissa server cond-mat/9611032.
\bibitem{Lu} J. P. Lu, Phys. Reb. B {\bf 49}, 5687 (1994).
\bibitem{Kotliar}G. Kotliar   and  A. E. Ruckenstein, Phys. Rev. 
Lett. {\bf 57}, 1362 (1986).
\bibitem{RFGK} R. Fr\'esard, G. Kotliar (unpublished)
\bibitem{Lavagna} M. Lavagna, Phys. Rev. B {\bf 41}, 142 (1990)
\bibitem{Okimoto} Y. Okimoto {\it et al}, Phys. Rev. B {\bf 51}, 9581 (1995).
\bibitem{Li91} 
T. Li, Y.S. Sun and P. W\"olfle, Z. Phys. B {\bf 82}, 369 (1991).
\bibitem{Doll} R. Fr\'esard and K. Doll,  Proc. of the
NATO ARW ``The Hubbard Model: Its Physics and Mathematical Physics'',
Ed. D. Baeriswil {\it et al},
San Sebastian (1993), Plenum (1995), p.385
\bibitem{Rozenberg} M. J. Rozenberg {\it et al}, Phys. Rev. Lett. {\bf 75},
105 (1995)
\bibitem{Moeller}B. M\"oller {\it et al}, J. Phys.: Cond. Matt. {\bf
5}, 4847 (1993).
\bibitem{Wurth} P. Wurth {\it et al} Sissa Preprint cond-mat/9512060.
\end{thebibliography}
\end{document}